\documentstyle[12pt]{article}

\advance\textheight by \topskip
\begin{document}

\thispagestyle{empty}
\setcounter{page}{0}

\begin{flushright}
{\large DFTT 4/93}\\
{\rm February 1993\hspace*{.5 truecm}}\\
\end{flushright}

\vspace*{\fill}

\begin{center}
{\Large \bf Hard Photon Pair Production
at LEP\footnote{ Work supported in part by Ministero
dell' Universit\`a e della Ricerca Scientifica.}}\\[2cm]
{\large Alessandro Ballestrero}\\[.3 cm]
{\it INFN, Sezione di Torino}\\[1cm]
{\large Ezio Maina, Stefano Moretti}\\[.3 cm]
{\it Dipartimento di Fisica Teorica, Universit\`a di Torino, Italy}\\
{\it and INFN, Sezione di Torino, Italy}\\[1cm]
\end{center}

\vspace*{\fill}

\begin{abstract}
{\normalsize
The production of photon pairs in $e^+e^-\rightarrow f \bar f \gamma\gamma$
processes
is studied using exact helicity amplitudes at tree level. Total
cross sections,
including initial state radiation effects, are given. They are
presented in the case of quarks
as a function of $y_{\rm cut}$ in the JADE algorithm.
In the case of leptons we use cuts similar to the
ones employed in a recent L3 analysis.
Masses of the final state fermions are taken
into account when appropriate. The cross section
$e^+e^-\rightarrow \tau^+ \tau^- \gamma\gamma$ is
about 10\% larger when the $\tau$ mass is neglected.
We obtain, with a different method,
results which are in good agreement with
L3 Montecarlo simulation.
}
\end{abstract}

\vspace*{\fill}

\newpage
\subsection*{Introduction}
Photon emission is of crucial importance in $e^+e^-$ collisions at energies
close to the $Z^0$ peak, and a good understanding of its properties is
necessary in order to test the Standard Model with high accuracy.
Most of the radiation is soft and/or collinear.
The cross section for events with hard, isolated photons is
much smaller, being approximately
reduced by a factor $\alpha_{em}$ for each emitted photon with respect to
the cross section with only soft radiation.
Hard photons are interesting because they probe the quark charges and
because they could be a signal for new particles
or for some kind of substructure of known particles
\cite{gamma1,gamma2}.
Not surprisingly, it has been shown \cite{stir} that the soft approximation is
inadequate to describe hard photon radiation, and that a matrix element
calculation is needed for this purpose.\par
The L3 Collaboration \cite{L3} has recently reported four events in which
two hard photons are emitted with a total invariant mass close to 60 GeV.
{}From a comparison with the YSF3 Montecarlo \cite{yfs3}, after full
detector simulation,
L3 estimates the probability for all four events to originate from
ordinary QED to be of the order of $10^{-3}$. The expected rate for the
production of a Standard Model Higgs of about 60 GeV, in association
with leptons, decaying in the two--photon channel, is about four orders
of magnitude smaller than the measured one.\par
The obvious thing to do in order to understand whether these peculiar events
represent the first manifestation of some new physics
is to examine all multi--photon events and compare them
with the available simulations in search of anomalies.\par
The L3 events have a typical $y_{\rm cut}$ in the $10^{-3}$ range,
and events with
$y_{\rm cut}$ as small as $10^{-4}$ would still survive all cuts.
This range has only been partially analyzed, with a simplified set
of cuts, in previous studies \cite{stir}.\par
In two recent papers \cite{last1,last2} we have shown
that at the $Z^0$ peak, when
masses are properly taken into account, cross sections involving
$c$ or $b$ quarks differ significantly from
the corresponding predictions obtained when masses
are neglected.
For final states with more than one photon the $b$-quark contribution is
severely reduced by the small electric charge of the $b$.
For $c$-quarks the mass effect is of the order of several
percent.
However, in order to test this result, $c$-quarks tagging
is required, and this substantially decreases an already small rate and
increases the sources of uncertainty.
These results suggest that a calculation for the
more easily accessible reaction
$e^+e^- \rightarrow \tau^+ \tau^- \gamma\gamma$, in which
the $\tau$ mass is not neglected, could give
significantly smaller predictions than those available so far.\par
The main purpose of this paper is to give a realistic estimate of the total
cross section and of some typical distribution
for $e^+e^- \rightarrow \tau^+ \tau^- \gamma\gamma$,
taking into account initial state radiation and the $\tau$ mass,
with a set of cuts which mimic as closely as possible the ones used in the L3
experiment. At the same time we study two--photon production
in association with jets, for which no anomalous event has been reported.
We also compute the cross section and some distributions for
$e^+e^- \rightarrow \mu^+ \mu^- \gamma\gamma$ in order to be able to
compare our calculations with L3 results and to determine if our
approach is adequate.\par
\subsection*{Calculation}
We consider processes without electrons in the final state,
which proceed through $e^+e^-$ annihilation into a photon or a $Z^0$.
Both contributions have been retained in all amplitudes.
In our calculation hard photon radiation is emitted only from
the final state fermions, while the radiation from the initial
electron legs is described using the structure function
approach \cite{nicrotre}.
This separation of final and initial state radiation makes sense since
the two terms are separately gauge invariant and we are only
interested in photons at relatively large angle with respect to the beam.
We have computed all matrix elements both in the formalism of \cite{ks,mana}
and in that of \cite{hz}.
The analytic expressions we have used, in the formalism of
\cite{ks,mana}, can be easily extracted from
Appendix B of \cite{last2}.
\par
The amplitudes have been checked for gauge invariance.
We have used  $M_Z=91.1$ GeV, $\Gamma_Z=2.5$ GeV,
$\sin^2 (\theta_W)=.23$, $\alpha_{em}= 1/128$ for the vertices
connected to the $Z^0, \gamma$ propagator and
$\alpha_{em}= 1/137$ everywhere else,
$m_\tau=1.78$ GeV and $m_b=5.$ GeV
in the numerical part of our work.\par
In what follows we neglect all hadronization effects, and
apply cuts at the partonic level.\par
\subsection*{Results}
The total cross sections for
$e^+e^-\rightarrow \mu^+ \mu^- \gamma\gamma$ and
$e^+e^-\rightarrow \tau^+ \tau^- \gamma\gamma$ at
$\sqrt s = M_Z=91.1$ GeV are given
in table I. For comparison we also give the cross section which
results neglecting the $\tau$ mass.
Only photons satisfying $m_{\gamma\gamma}>2.5$ GeV,
$|p_\gamma|>1$ GeV and $|cos\theta_\gamma|<0.9$
are accepted. For $\mu$'s we require
$|p_\mu|>3$ GeV,
$|cos\theta_\mu|<0.81$ and $\theta_{\gamma\mu}>5^{\circ}$.
For $\tau$'s and for $\ell_0$'s, the reference massless leptons
we compare the  $\tau$'s with, the assumed cuts are
$|cos\theta_\ell|<0.74$ and
$\theta_{\gamma\ell}>15^{\circ}$.\par
In fig.1 we present the total cross sections for
$e^+e^-\rightarrow q \bar q \gamma\gamma$ for $u$, $d$ and
$b$-quarks separately
and then for the sum over five flavors as a function of $y_{\rm cut}$.
Only for $b$-quarks the mass is taken into account.
For photon pairs in association with jets
we have required
$\theta_{\gamma q}>15^{\circ}$ but have not imposed any
cut on the angles with respect to the beam. All particle pairs
must have $y > y_{\rm cut}$, where $y$ is the usual JADE
variable \cite{jade}.
\par
In fig.2 we present the energy distribution of the most energetic photon
in $e^+e^-\rightarrow \mu^+ \mu^- \gamma\gamma$,
$e^+e^-\rightarrow \tau^+ \tau^- \gamma\gamma$ and
$e^+e^-\rightarrow \ell_0^+ \ell_0^- \gamma\gamma$
with $\ell_0$ massless.
In fig.3 we give the two--photon mass distribution
in $e^+e^-\rightarrow \tau^+ \tau^- \gamma\gamma$ and again compare
it with the results for a massless lepton with the same choice of cuts.
Finally in fig.4 we show the two--photon invariant mass distribution
for
$e^+e^-\rightarrow \mu^+ \mu^- \gamma\gamma$,
$e^+e^-\rightarrow \tau^+ \tau^- \gamma\gamma$ and
$e^+e^-\rightarrow q \bar q \gamma\gamma$  summed over all flavors.\par
In fig.2 and 4 the results are deliberately presented in a way that
allows an easy comparison with L3 data.\par
In order to discuss the consistency of our results with
L3 Montecarlo simulations we proceed as follows.
{}From the fraction of muon events in the Montecarlo sample
which survives
all cuts, an efficiency of about 56\% can be inferred for the L3
detector. The main factors contributing to the
efficiency are the geometrical acceptance and
the effect of the dead region between detector
modules in the R-$\phi$ plane within
the fiducial volume $|cos\theta_\mu|<0.81$.
The first contribution can
be easily extracted from the ratio of the integrated
Born cross section
for for $e^+e^-\rightarrow \mu^+ \mu^-$, in the presence
of the angular and momentum cuts and in their absence,
and turns out to be  about 74\%.
The additional effect related to inefficiencies
within the fiducial volume can therefore be estimated at about 76\%.
\par
L3 data include runs from 1990 through July 1992
at various energies around the $Z^0$ peak.
An equivalent peak--luminosity of about 22 $pb^{-1}$
can be derived, dividing
the total sample of 950,000 $Z^0$ by the peak-cross section
of about 43 $nb$, including initial state radiation.
This luminosity and the 76\% efficiency in detecting muons
can then be used to translate
the cross sections we present, which already include
geometrical acceptance, in a prediction for the number of events.
As an example we would expect 50 events with two photons and a
$\mu^+\mu^-$ pair in the sample.
In order to compare this result with the L3 Montecarlo simulation it is
necessary to correct for the cut in the invariant mass of the photon pair
which is present in our analysis and is not implemented by L3.
{}From the $m_{\gamma\gamma}$ distribution in fig.6 of \cite{L3}
one can read that the first mass bin, which exactly corresponds to
our excluded region, contains 20\% of the Montecarlo events.
Therefore our prediction of 50 events has to be compared with
45 events predicted by L3 in the same mass range.
This agreement at the 10\% level is, in our opinion,
quite satisfactory.
The integrated cross section for
$m_{\gamma\gamma} > 50$ GeV is $3.5\times10^{-2}$ $pb$ for
$\mu^+ \mu^- \gamma\gamma$, corresponding to .6 events.\par
The consistency of the results obtained in the two approaches,
for photon pairs in association with leptons,
can be better appreciated looking at the shape of
the mass distribution in fig.4 and in fig.6 of \cite{L3}.
There is also good agreement, for large photon energies,
for the shape of the energy distribution of the most energetic photon
shown in
fig.2 and in fig.2a of \cite{L3}. The low energy peak,
present in fig.2a of \cite{L3},
is strongly reduced in our distribution due to the additional
minimum invariant mass requirement.\par
On one hand we interpret these results as a confirmation of the correctness
of L3 Montecarlo simulation, on the other hand they give us confidence
in our predictions.\par
{}From table I we see that the predicted cross section for
$e^+e^-\rightarrow \tau^+ \tau^- \gamma\gamma$ is about 9\% smaller than it
would be for a massless lepton. Fig.2 and 3 show that the effect
of the $\tau$ mass is more
pronounced for large photon energies and for large invariant mass of the photon
pair.
As expected, the total cross section is dominated by
low energy photons and low mass pairs.\par
The cross sections for photon pairs in association with jets,
with the assumed set of cuts and $y_{\rm cut}$ = $5\times 10^{-3}$,
is about 1 $pb$ corresponding to
22 events. Since the mentioned cuts are similar to those employed in
\cite{gamma2} we expect a number of such events to be present in the data
sample of each LEP experiment.\par
As a byproduct of the calculations presented here, we have verified that
the ratio of distributions for massive particles to those for massless ones
is not modified when initial state radiation is included.
Therefore, also in this more realistic case, all conclusions drawn
in \cite{last1,last2} remain valid.
Obviously all distributions have to be scaled down by approximately 30\%.
\subsection*{Conclusions}
We have produced the total
cross section and some typical distributions
for $e^+e^- \rightarrow \tau^+ \tau^- \gamma\gamma$,
taking into account initial state radiation and the $\tau$ mass,
with a set of cuts which mimic as closely as possible the ones used in the L3
experiment. The total cross section is 9\% smaller than it would be for a
massless lepton and the discrepancy increases for larger photon energies and
two--photon invariant masses.
Photon pair production
in association with jets has also been computed and we believe it would be
important to look for such events.
We have studied $e^+e^- \rightarrow \mu^+ \mu^- \gamma\gamma$
obtaining a total rate and distributions in good agreement with L3 results.

\subsection*{Acknowledgements}
We gratefully acknowledge the kind collaboration of F. Ferroni
and J.M. Qian in understanding various aspects of the L3
analysis.

\newpage

\subsection*{Table Captions}
\begin{description}
\item[table I.] Cross sections, cuts and parameters
for $e^+e^-\rightarrow\ell^+\ell^-\gamma\gamma$ with
$\ell=\mu,\tau,\ell_0$,
where $\ell_0$ is a massless lepton.
Errors are as calculated by VEGAS \cite{vegas}.
\end{description}

\vspace*{\fill}

\subsection*{Figure Captions}
\begin{description}
\item[fig.1.] Cross sections for
$e^+e^-\rightarrow u \bar u \gamma\gamma$,
$e^+e^-\rightarrow d \bar d \gamma\gamma$,
$e^+e^-\rightarrow b \bar b \gamma\gamma$ and
for the sum over five flavors
as a function of $y_{\rm cut}$ at $\sqrt{s} = 91.1$ GeV.
The cuts used for photon
pairs in association with jets are $y_{\rm cut} > 5\times 10^{-3}$,
$\theta_{\gamma q}>15^{\circ}$ and $|p_\gamma|>1$ GeV.

\item[fig.2.]  Energy distribution of the most energetic
photon for
$e^+e^-\rightarrow \mu^+ \mu^- \gamma\gamma$,
$e^+e^-\rightarrow \ell_0^+ \ell_0^- \gamma\gamma$
and for $e^+e^-\rightarrow \tau^+ \tau^-  \gamma\gamma$,
where $\ell_0$ is massless, at $\sqrt{s} = 91.1$ GeV.
Cuts as in table I.

\item[fig.3.]  Two--photon invariant mass distributions
for $e^+e^-\rightarrow \ell_0^+ \ell_0^- \gamma\gamma$
and for $e^+e^-\rightarrow \tau^+ \tau^-  \gamma\gamma$,
where $\ell_0$ is massless, at $\sqrt{s} = 91.1$ GeV.
Cuts as in table I.

\item[fig.4.]  Two--photon invariant mass distributions for
$e^+e^-\rightarrow \mu^+ \mu^- \gamma\gamma$,
$e^+e^-\rightarrow \tau^+ \tau^- \gamma\gamma$ and
for $e^+e^-\rightarrow q \bar q \gamma\gamma$,
with $y_{\rm cut}$ = $5\times 10^{-3}$ and summed over
five flavors, at $\sqrt{s} = 91.1$ GeV.
Cuts as in table I for leptons. The cuts used for photon
pairs in association with jets are $y_{\rm cut} > 5\times 10^{-3}$,
$\theta_{\gamma q}>15^{\circ}$ and $|p_\gamma|>1$ GeV.
\end{description}

\vspace*{\fill}

\newpage
$$\vbox{\tabskip=0pt \offinterlineskip
\halign to400pt{\strut#& \vrule#\tabskip=1.em plus2.0em& \hfil#&
\vrule#& \hfil#&
\vrule#& \hfil#&
\vrule#\tabskip=0pt\cr \noalign{\hrule}
&  && \omit && \omit && \cr
&  && \omit &$\sigma$ (pb)~~~~~~~~~~~~& \omit && \cr
&  && \omit && \omit && \cr
\noalign{\hrule}
&  && && && \cr
&  &$e^+e^-\rightarrow\mu^+\mu^-\gamma\gamma$~~~~&
   &$e^+e^-\rightarrow\tau^+\tau^-\gamma\gamma$~&
   &$e^+e^-\rightarrow\ell_0^+\ell_0^-\gamma\gamma$~& \cr
&  && && && \cr
\noalign{\hrule}
&  && && && \cr
&  &$(4.093\pm0.016)^{*}$~~~~~&
   &$(1.4746\pm0.0038)^{*}$&
   &$(1.6169\pm0.0052)^{*}$& \cr
&  && && && \cr
&  &$(3.025\pm0.015)^{\dag}$~~~~~&
   &$(1.0854\pm0.0035)^{\dag}$&
   &$(1.1913\pm0.0048)^{\dag}$& \cr
&  && && && \cr
\noalign{\hrule}
&  && && \omit && \cr
&  &$|p_\ell|>3$ GeV~~~~~~~~& && \omit && \cr
&  && &$|cos\theta_\ell|<0.74$~~& \omit
   &$\theta_{\gamma\ell}>15^{\circ}$~~~~~& \cr
&  &$|cos\theta_\ell|<0.81$~~$\theta_{\gamma\ell}>5^{\circ}$&
   && \omit && \cr
&  && && \omit && \cr
\noalign{\hrule}
&  && \omit && \omit && \cr
&  &$m_{\gamma\gamma}>2.5$ GeV~~~~~& \omit
   &$|p_\gamma|>1$ GeV~~~& \omit
   &$|cos\theta_\gamma|<0.9$~~~~& \cr
&  && \omit && \omit && \cr
\noalign{\hrule}
&  && \omit && \omit && \cr
&  &$\sqrt s=M_Z=91.1$ GeV~& \omit
   &$\Gamma_Z=2.5$ GeV~~~& \omit
   &$\sin^2\theta_W=0.23~~~$& \cr
&  && \omit && \omit && \cr
\noalign{\hrule}
\noalign{\smallskip}
& \multispan7 $^{*}$ Without ISR; $^{\dag}$ With ISR.
\hfil\cr}}$$
\begin{center}
Table I
\end{center}

\end{document}